\documentclass{article}
\usepackage{spconf,amsmath,graphicx}

\usepackage[nolist,nohyperlinks]{acronym}
\usepackage{comment}
\usepackage{hhline}
\usepackage{bm}
\usepackage{subcaption}
\usepackage[hyphens]{url}

\newcommand{\squeezeup}{\vspace{-2.5mm}}

\acrodef{PERL}{Perceptual Ensemble Regularization Loss}
\acrodef{TNN}{Transformer Neural Network}
\acrodef{ASR}{Automatic Speech Recognition}
\acrodef{GAN}{Generative Adversarial Network}
\acrodef{SOTA}{state-of-the-art}
\acrodef{DFL}{Deep Feature Loss}
\acrodef{MTL}{Multi-Task Learning}
\acrodef{SIDE}{Single-Image Depth Estimation}
\acrodef{TTS}{Text-to-Speech}
\acrodef{DNN}{Deep Neural Network}
\acrodef{MOS}{Mean Opinion Score}
\acrodef{SESQA}{Semi-supervised Speech Quality Assessment}
\acrodef{PMSQE}{Perceptual Metric for Speech Quality Evaluation}
\acrodef{FFNs}{Feed-Forward Networks}
\acrodef{FFN}{Feed-Forward Network}
\acrodef{PE}{Position Encoding}
\acrodef{BN}{Batch Normalization}
\acrodef{PANNs}{Pre-trained Audio Neural Networks}
\acrodef{PANN}{Pre-trained Audio Neural Network}
\acrodef{DCASE}{Detection and Classiﬁcation of Acoustic Scenes and Events}
\acrodef{SER}{Sound Event Recognition}
\acrodef{LMFB}{Log Mel-FilterBank}
\acrodef{CNN}{Convolutional Neural Network}
\acrodef{FMS}{Feature Map Scaling}
\acrodef{SE}{Squeeze-and-Excitation}
\acrodef{TTA}{Test-Time Augmentation}
\acrodef{DS2}{Deep Speech 2}
\acrodef{CTC}{Connectionist Temporal Classification}
\acrodef{PASE}{Problem-Agnostic Speech Encoder}
\acrodef{SDR}{Signal-to-Distortion Ratio}
\acrodef{PESQ}{Perceptual Evaluation of Speech Quality}
\acrodef{QRNN}{Quasi-Recurrent Neural Network}
\acrodef{VAD}{Voice Activity Detection}
\acrodef{PReLU}{Parameteric ReLU}
\acrodef{NCP}{Noise Corruption Process}
\acrodef{SER}{Speech Emotion Recognition}
\acrodef{GRU}{Gated Recurrent Units}
\acrodef{RNN}{Recurrent Neural Network}
\acrodef{BLSTM}{Bi-directional Long Short Term Memory}
\acrodef{TDNN}{Time-Delay Neural Network}
\acrodef{SA}{Signal Approximation}
\acrodef{STFT}{Short-Time Fourier Transform}
\acrodef{JND}{Just Noticeable Difference}
\acrodef{JNDs}{Just Noticeable Differences}
\acrodef{TDR}{Time-Domain Reconstruction}

\title{Perceptual Loss based Speech Denoising with\\an ensemble of Audio Pattern Recognition and Self-Supervised Models}

\name{Saurabh Kataria, Jes\'us Villalba, Najim Dehak}
\address{\{skatari1,jvillal7,ndehak3\}@jhu.edu\\
Center for Language and Speech Processing, Johns Hopkins University, Baltimore, MD, USA}
\begin{document}
\ninept
\maketitle
\begin{abstract}
Deep learning based speech denoising still suffers from the challenge of improving perceptual quality of enhanced signals.
We introduce a generalized framework called Perceptual Ensemble Regularization Loss (PERL) built on the idea of \emph{perceptual losses}.
Perceptual loss discourages distortion to certain speech properties and we analyze it using six large-scale pre-trained models: speaker classification, acoustic model, speaker embedding, emotion classification, and two self-supervised speech encoders (PASE+, wav2vec 2.0).
We first build a strong baseline (w/o PERL) using Conformer Transformer Networks on the popular enhancement benchmark called VCTK-DEMAND.
Using auxiliary models one at a time, we find acoustic event and self-supervised model PASE+ to be most effective.
Our best model (PERL-AE) only uses acoustic event model (utilizing AudioSet) to outperform state-of-the-art methods on major perceptual metrics.
To explore if denoising can leverage full framework, we use all networks but find that our seven-loss formulation suffers from the challenges of Multi-Task Learning.
Finally, we report a critical observation that state-of-the-art Multi-Task weight learning methods cannot outperform hand tuning, perhaps due to challenges of domain mismatch and \emph{weak complementarity} of losses.
\end{abstract}
\section{Introduction}
There is a growing focus on the perceptual and intelligibility quality of enhanced signals obtained from \ac{DNN} based speech enhancement systems.
Quantifying them through human \ac{MOS} is highly expensive and prone to error.
Proxy objective metrics (reference/non-reference based) are used in enhancement like \ac{PESQ} but they are hard to improve upon.
In \cite{germain2018speech}, authors introduced \emph{perceptual loss} (or \ac{DFL}) to speech denoising problem.
This improves perceptual quality of audio and is successfully applied using \ac{GAN}~\cite{su2020hifi} and also for \emph{task-specific} enhancement~\cite{kataria2020analysis}.

Alternatively, various differentiable metrics are proposed like \ac{PMSQE}~\cite{martin2018deep}, \ac{SESQA}~\cite{serra2020sesqa}, and a metric based on \ac{JNDs}~\cite{manocha2020differentiable}.
In \cite{serra2020sesqa}, authors proposed \ac{SESQA} which is an eight-loss \ac{MTL} based semi-supervised framework for automated speech quality assessment.
In addition to predicting \ac{MOS}, they define various auxiliary tasks including predicting JND, pairwise comparison, degradation strength, etc.
\cite{manocha2020differentiable} took a different approach by constructing a large corpus based on (binary) human preference in audio pairs.
By capturing JND, they are able to train a differentiable metric well correlated with human \ac{MOS} rating.

Our aim is to improve perceptual and intelligibility metrics using \emph{perceptual loss} on small-scale supervised speech denoising.
To compensate for low resource training data, we investigate if large-scale pre-trained speech models can be utilized.
Such models, when used frozen and in ensemble, can regularize enhancement training to minimize distortions to various speech properties.
Furthermore, we explore if self-supervised models can help preserve speech representations since they are trained on generic \emph{pretext tasks}.
Our idea faces challenges of domain mismatch and weak complementarity of losses, which we explore in-depth with various \ac{MTL} methods.

Our contributions are as follows.
First, we establish a strong baseline based on Conformer Transformer~\cite{gulati2020conformer} with a simple $l_1$ objective.
Second, we propose a rich framework called \ac{PERL} which leverages six open-source large-scale pre-trained networks to analyze speech denoising with perceptual losses.
Third, we rank various speech tasks in term of their effectiveness in PERL and, for the first time, we demonstrate the utility of self-supervised representations for perceptual loss training.
Fourth, we show the importance of $l_1$ enhancement loss and all Conformer components to achieve \ac{SOTA} performance.
Fifth, using all seven losses of PERL, we make an important finding that \ac{SOTA} multi-task weight learning methods cannot outperform hand-tuned weights.

\begin{figure*}[ht]
    \centering
    \includegraphics[scale=0.58]{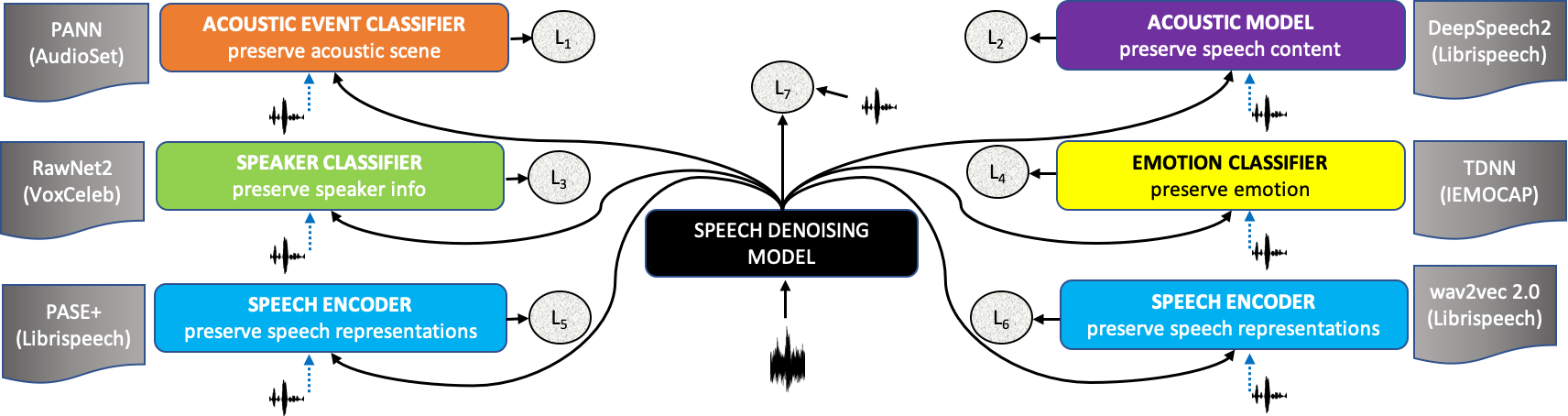}
    \caption{Illustration of supervised framework called Perceptual Ensemble Regularization Loss (PERL) using seven losses: $L_1,\dots,L_7$. To extract perceptual losses, partial forward pass of enhanced and reference clean signals is done as illustrated by using temporal signals.}
    \squeezeup
    \label{fig:perl}
\end{figure*}

\section{Speech Denoising with Perceptual Ensemble Regularization Loss (PERL)}
\label{sec:perl}
We consider a simple signal model or \ac{NCP}: $y(t) = x(t) + n(t)$, where $x(t)$ is clean (time-domain) signal, $n(t)$ is noise signal, and $y(t)$ is the resultant noisy signal.
In general, $x(t)$ suffers distortions to its acoustic event, speaker characteristics, and acoustic content.
In \ac{DNN} based speech enhancement, a simple $l_1$ loss is not able to restore such distortions~\cite{kataria2020analysis}.
Moreover, enhancement may introduce such distortions~\cite{kataria2020analysis}.
To counter this, \cite{germain2018speech,kataria2020analysis} used \emph{perceptual loss} or \emph{deep feature loss}.
Such loss compares enhanced signals with reference clean signals in the activation space of a pre-trained auxiliary network (trained for task $\mathcal{T}$).
Depending on the choice of the auxiliary network, we accomplish speech enhancement with minimal distortions to representations required to solve task $\mathcal{T}$.
For e.g., \cite{kataria2020analysis} did \ac{DFL} based speech enhancement while minimizing distortions to speaker identities.

To avoid distortions to various speech characteristics during enhancement and/or \ac{NCP}, we propose \ac{PERL}: an ensemble of varied types of pre-trained (and frozen) speech models for perceptual loss training.
Specifically, we use acoustic event classification, speaker embedding, acoustic model, and \ac{SER}.
We also incorporate self-supervised speech encoders since they model general speech representations~\cite{baevski2020wav2vec}.
Our scheme is illustrated in Fig. \ref{fig:perl}.
When using all components, loss function is
\begin{equation}
\begin{aligned}
    \mathcal{L} = \hspace{1mm}&\lambda_{\text{event}}\mathcal{L}_{\text{event},n_1} + \lambda_{\text{acoustic}}\mathcal{L}_{\text{acoustic},n_2} + \lambda_{\text{speaker}}\mathcal{L}_{\text{speaker},n_3}\\ +& \lambda_{\text{emotion}}\mathcal{L}_{\text{emotion},n_4} + \lambda_{\text{pase}}\mathcal{L}_{\text{pase},n_5} + \lambda_{\text{wav2vec}}\mathcal{L}_{\text{wav2vec},n_6}\\
    +&\lambda_{l_1}\mathcal{L}_{l_1},
\end{aligned}
\end{equation}
where the last term refers to the simple \ac{STFT} feature-domain $l_1$ enhancement loss and $n_i$ refers to the number of internal layers of the corresponding auxiliary network used for deriving perceptual loss.
The perceptual losses derived from $i$-th network are summed and divided by $n_i$.
We term \ac{PERL} as a \emph{regularization loss} since the auxiliary networks are frozen and only help constrain the output space.
Note that this loss is different from usual formulations of transfer learning, knowledge distillation, and even multi-task learning.
\cite{groenendijk2020multi} appropriately terms such setting as \emph{single-task multi-loss}.
Our framework is flexible and, instead of speech denoising, it can be used for voice conversion and domain adaptation as well.
In terms of framework richness, our work is close to \ac{SESQA}~\cite{serra2020sesqa} and PASE+~\cite{ravanelli2020multi}.

There are some inherent challenges with \ac{PERL}.
One, since the auxiliary networks are trained on different features and domain, there can be significant domain mismatch which hinder training.
To avoid this, we prefer to use initial layers since they tend to be more generic.
Two, our test set is not equipped with labels for all auxiliary tasks and hence evaluation is restricted to enhancement metrics only.

\section{Model Descriptions}
\subsection{Baseline Conformer Transformer Enhancement Network}
Following the success of Transformer-based modeling of speech features~\cite{gulati2020conformer,kim2020t}, we choose \emph{convolution-augmented Transformer} or Conformer \cite{gulati2020conformer} for the denoising network.
For modeling long-term and short-term patterns, it relies on self-attention mechanism and specially designed \emph{convolution modules} respectively.
Moreover, it combines the power of relative \ac{PE} scheme and Macaron-style half-step \ac{FFNs}~\cite{gulati2020conformer}.
We additionally include Squeeze-and-Excitation~\cite{hu2018squeeze} module (squeeze factor of 8) after the \emph{1D Depthwise Convolution} inside the \emph{convolution module} of Conformer.
To avoid down-sampling in time, for the first layer of our Conformer, we use a \ac{BN} layer followed by a linear transformation instead of the usual convolutional sub-sampling layer.
The attention dimension of the network is 240, the number of conformer blocks are 4, and total parameters are 10M.
The denoising network predicts a mask which is then multiplied with the noisy spectra to predict the clean spectra.

\subsection{Brief Overview of Classification Models}
\underline{\textbf{Acoustic Event Classification}}: In \cite{kong2019panns}, authors train a large-scale audio event classification network called \ac{PANN} and prove its transferability to six audio pattern recognition tasks.
It is trained with 1.9M audio clips from AudioSet~\cite{gemmeke2017audio} (5000hrs, 527 sound classes).
We choose the wide-band 14-layer version
which has 81M parameters.
They also address data imbalance problem of AudioSet and do online data augmentation with MixUp and SpecAugment~\cite{park2019specaugment}.
We choose $n_\text{event}=4$ by using the first four convolution blocks (or first eight convolution layers).

\underline{\textbf{Acoustic Model}}: \ac{DS2} is a multi-lingual end-to-end \ac{ASR} model proposed in \cite{amodei2016deep}.
We choose the English model trained on LibriSpeech~\cite{panayotov2015librispeech}
consisting of 13M parameters.
The architecture follows the \ac{RNN}-\ac{CTC} style.
It has a convolutional front-end which is followed by five \ac{BLSTM} layers and final classification/CTC layer for English graphemes.
We choose $n_{\text{acoustic}}=3$ by using the output of the first two convolutional layers and first RNN layer for DFL.

\underline{\textbf{Speaker Embedding}}: In \cite{jung2020improved}, authors propose RawNet2.
It is a time-domain speaker embedding network trained with Voxceleb~\cite{jung2020improved}, which contains over 1M utterances from 6112 speakers.
The architecture
followed is a \ac{CNN} made of six residual blocks followed by \ac{GRU}.
The first layer is a sinc-convolution layer of SincNet~\cite{ravanelli2018speaker}.
Total number of parameters are 87M.
We choose $n_\text{speaker}=3$ by using the output of the first sinc layer and the next two convolutional blocks.

\underline{\textbf{Speech Emotion Classification}}: \cite{sarma2018emotion} trains a time-domain emotion identification network on a 12hr emotion corpus called IEMOCAP.
The architecture follows a \ac{TDNN}-LSTM style and has 1M parameters.
The first layer is a 1-D convolution based data dependent layer.
It also consists of five LSTM layers followed by four \ac{TDNN} layers and a time-restricted self-attention based pooling of embeddings.
Each frame of speech is classified in four emotions: happy, sad, neutral, and angry.
We train this network
using the same training scheme used for Conformer as described in Sec. \ref{sec:exp}.
We choose $n_{\text{emotion}}=3$ by using the output of the first three layers.

\subsection{Brief Overview of Self-Supervised Speech Encoder Models}
\underline{\textbf{PASE+}}: PASE+~\cite{ravanelli2020multi} is a self-supervised speech encoder trained with 12 self-supervised tasks.
For \emph{regression tasks}, the model learns to predict various known speech transformations.
For \emph{binary tasks}, a contrastive loss tries to bring close representations of speech chunks belonging to same utterance while pushing apart representations of chunks belonging to different utterances.
PASE+ can model speech content and speaker properties, which are robust to noise perturbations and have good transferability.
The architecture has a convolutional front-end with a \ac{QRNN} backbone.
It is trained with 50hrs of Librispeech~\cite{panayotov2015librispeech} and number of trainable parameters are 8M.
We interchangeably refer to this model as PASE.
We use $n_{\text{pase}}=6$ by using the output of the first six convolutional blocks.

\underline{\textbf{wav2vec 2.0}}
In \cite{baevski2020wav2vec}, authors propose wav2vec 2.0, a self-supervised model which learns generic speech representations by solving a contrastive loss in the (quantized) latent space using a masked Transformer.
It is trained on 960hrs of LibriSpeech without transcriptions.
The architecture
 consists of 12 transformer blocks, model dimension of 768, inner dimension (\ac{FFN}) of 3,072, eight attention heads, and 96M parameters.
We interchangeably refer to this model as wav2vec.
We choose $n_{\text{wav2vec}}=5$ by using the output of the convolutional front-end and the next four Transformer blocks.

\subsection{Brief Overview of Multi-Task Learning Methods}
We use four \ac{MTL} methods off-the-shelf whose some aspects are as follows.
\emph{Uncertainty Weighting}~\cite{kendall2018multi} uses the notion of an inherent \emph{task-dependent uncertainty} or \emph{homoscedastic uncertainty}.
Using a Gaussian likelihood formulation, it defines a loss function consisting of learnable variance parameters for each loss.
Hence, it accounts for the dynamic range of loss terms in a principled way.
\emph{Coefficient of Variation}~\cite{groenendijk2020multi} states that a loss term is satisfied when its variance has decreased to zero.
Hence, it assigns weight for a loss term as its coefficient of variation, i.e., standard deviation divided by mean.
This quantity can compensate for the different dynamic range of loss terms.
For mean estimate, a robust formulation of the current loss is used.
For variance estimate, an online statistics tracking algorithm is used.
Finally, all weights are normalized to make sum equals to one.
\emph{Dynamic Weight Averaging}~\cite{liu2019end} keeps track of instantaneous ratio of change of loss value.
Using that and a temperature parameter $T$ (fixed to 2), it assigns weight to each task via a softmax classification.
\emph{GradCosine}~\cite{du2018adapting} treats main and auxiliary loss separately.
To update the shared parameters (in our case, full denoising network), cosine similarity of gradient w.r.t. auxiliary loss is checked against the gradient w.r.t. main loss.
When auxiliary gradients align, they are used to update shared parameters, otherwise they are rejected.

\section{Experimental Setup}
\label{sec:exp}
We evaluate speech denoising on VCTK-DEMAND~\cite{valentini2016investigating}.
It is a 16KHz small corpus with fixed training (10hr, 30 speakers) and validation data (30m, 2 speakers).
We use the predicted clean spectra to compute \ac{SA} loss or, simply, the $l_1$ loss (in STFT domain).
We experimented with several popular loss functions but found $l_1$ loss to be the best and the most resilient to change in experimental setup.
For compatibility with auxiliary networks, we do appropriate on-the-fly transformations.
To convert to time-domain, we re-use the phase of the noisy signal.
Due to the large number of experiments, instead of subjective evaluation, we aim to improve upon standardized perceptual and intelligibility metrics like PESQ, CSIG, CBAK, COVL, and STOI.
PESQ $\in [-0.5,4.5]$, STOI (in \%) $\in [0,100]$, while other metrics $\in [0,5]$.
Our seven-network based framework is memory intensive, and hence, we do gradient accumulation.
We train for 10 epochs using Adam optimizer, batch size of 32, a simple learning rate scheduler, and an initial learning rate of 0.00075.
Best scores on validation data are reported.

\begin{table}[t]
\centering
\caption{Comparison of perceptual losses of various speech models with simple $l_1$ loss. $n_p$ refers to \#parameters of auxiliary network.}
\resizebox{0.48\textwidth}{!}{
\begin{tabular}{|c|c|c|c|c|c|c|}
\hline
  &  $n_p$ & PESQ & CSIG & CBAK & COVL & STOI\\ \hline
$l_1$           &  -  & 3.01 & 4.27 & 3.48 & 3.65 &   \textbf{94.9} \\ \hhline{|=|=|=|=|=|=|=|}
AcousticEvent   & 81& \textbf{3.09} & \textbf{4.38} & \textbf{3.43} & \textbf{3.75} &    94.8\\ \hline
AcousticModel   & 87& 2.97 & 4.15 & 3.43 & 3.55 & 94.6 \\ \hline
SpeakerEmbedding &13 & 2.86 & 3.63 & 3.32 & 3.23 & 94.5 \\ \hline
SpeechEmotion   & 1 & 2.41 & 3.49 & 3.07 & 2.93 & 92.8 \\ \hline
PASE        &   8  & 3.04 & 4.23 & 3.42 & 3.63 & 94.7 \\ \hline
wav2vec       &  95 & 2.93 & 3.67 & 3.33 & 3.29 & 93.9 \\ \hline
\end{tabular}
}
\label{tab:base}
\end{table}

\section{Results}
\subsection{Comparison of perceptual losses of various speech models}
In Table \ref{tab:base}, we first note that using a simple $l_1$ loss in STFT domain with Conformers, we get close to \ac{SOTA} performance (Table \ref{tab:sota}).
In \cite{germain2018speech}, authors used acoustic event classifier to achieve \ac{PESQ} score of 2.57.
Using Conformers and AudioSet based auxiliary network, we drastically improve that score (3.09).
We find other speech models also give competitive performance except for the emotion model, perhaps due to its low resource training data.
Here, $n_p$ refers to the number of parameters in auxiliary network (in M).
PASE+ is the best model considering performance as well as parameter efficiency.
We experimented with various popular loss functions based on the ideas of multi-resolution spectra, \ac{TDR}, \ac{SDR}, etc. but found them much inferior to the $l_1$ loss.
It is important to state that we expect acoustic event loss to be the best since during \emph{noise corruption process} (and consequently during enhancement), for a simple test set like ours, \emph{acoustic event information of audio is adversely affected the most}.
However, its comparison with other speech models is novel.

\begin{table}[ht]
\centering
\caption{Effect of adding \emph{perceptual losses} (individually and in certain combinations) to $l_1$ loss. Last row uses all seven loss terms.}
\resizebox{0.48\textwidth}{!}{
\begin{tabular}{|l|l|l|l|l|l|}
\hline
                                                                                                        & PESQ & CSIG & CBAK & COVL & STOI \\ \hline
$l_1$                                                                                                      & 3.01 & 4.27 & 3.48 & 3.65 &    94.9  \\ \hhline{|=|=|=|=|=|=|}
$l_1$+AcousticEvent                                                                                        & \textbf{3.17} &  \textbf{4.43} & \textbf{3.53}  &  \textbf{3.83}    &   \textbf{95.0}         \\ \hline
$l_1$+AcousticModel                                                                                        & 2.97 &   4.15   &  3.43    &   3.55   &    94.6  \\ \hline
$l_1$+SpkEmbedding                                                                                         & 2.91 & 4.18     &  3.39    &  3.55    &    94.4  \\ \hline
$l_1$+SpkEmotion                                                                                           & 2.86 &     4.12 &  3.38    &  3.5    &     94.2 \\ \hline
$l_1$+PASE                                                                                                 & 3.03 &  4.24    &  3.43    &  3.63    &    94.8  \\ \hline
$l_1$+wav2vec                                                                                              & 2.92 & 4.16     &  3.37    &   3.54   &    94.5  \\ \hline
\begin{tabular}[c]{@{}l@{}}$l_1$+AcousticEvent\\ +AcousticModel\\ +SpkEmbedding\\ +SpkEmotion\end{tabular} & 2.88    &  4.15    &  3.40    &    3.52  &     94.5 \\ \hline
$l_1$+PASE+wav2vec                                                                                         & 2.95 &   4.2   &   3.4   &     3.58 &  94.6    \\ \hline
\begin{tabular}[c]{@{}l@{}}$l_1$+PASE\\ +AcousticEvent\\ +SpkEmbedding\end{tabular}                        & 3.05 &  4.31      &    3.5  &  3.69    &  94.8    \\ \hhline{|=|=|=|=|=|=|}
\begin{tabular}[c]{@{}l@{}}$l_1$+PASE\\ +AcousticEvent\end{tabular}                        & 3.04 &  4.37      &    3.47  &  3.72    &  94.9    \\ \hhline{|=|=|=|=|=|=|}
$l_1$+ALL (\textbf{PERL})                                                                                           & 2.89 &   4.11   &  3.41    &  3.5    & 94.6     \\ \hline
\end{tabular}
}
\label{tab:comb}
\end{table}

\subsection{Combining $l_1$ enhancement loss with perceptual losses}
Previous work~\cite{germain2018speech} did not investigate the benefit of combining feature-domain enhancement loss with perceptual losses.
In Table \ref{tab:comb}, we combine $l_1$ loss with six auxiliary losses individually as well in certain combinations.
Hand-tuning of loss terms gave $(\lambda_{\text{event}}, \lambda_{\text{acoustic}}, \lambda_{\text{speaker}}, \lambda_{\text{emotion}}, \lambda_{\text{pase}}, \lambda_{\text{wav2vec}}, \lambda_{l_1})$ = (5e-03, 1e-04, 1.25e-04, 4e-05, 1.7e-04, 3.5e-05, 1.1e-01).
Combining $l_1$ with acoustic event (PERL-AE) yields better results than baseline and outperforms \ac{SOTA} models (Table \ref{tab:sota}).
Note that we can further improve by predicting phase too by using a time-domain model like DEMUCS \cite{defossez2020real}.
For all cases, combining $l_1$ loss results in better performance than without it (Refer Table \ref{tab:base}).
Second last row in Table \ref{tab:comb} shows good performance by combining best three models: PASE, acoustic event, and speaker embedding.
This suggests that if we use such networks with more data and less domain mismatch w.r.t. enhancement training data, further improvements can be observed.
The final row uses all losses and, hence, is termed PERL.
Using full framework, we get inferior results w.r.t. baseline.
This is perhaps due to sub-optimal choice of loss weights (selected via greedy hand-tuning) or simply due to detrimental nature of some loss terms (investigated in Sec. \ref{sec:multi}).

To emphasize the role of Conformer architecture, we do an ablation study similar to \cite{gulati2020conformer}.
In Table \ref{tab:ablation}, we progressively (1) replace SWISH activation by ReLU; (2) remove \emph{convolution module} inside Conformer blocks; (3) replace Macaron-style FFN pairs with a single FFN layer; (4) replace relative position embedding with absolute.
The trend of performance degradation shows that all components of Conformer are important, especially the \emph{convolution module}.

\begin{table}[t]
\centering
\caption{Comparison of the best denoising system (PERL-AE or $l_1$+AcousticEvent) with the state-of-the-art methods}
\resizebox{0.48\textwidth}{!}{
\begin{tabular}{|c|c|c|c|c|c|}
\hline
          & PESQ & CSIG & CBAK & COVL & STOI \\ \hline
Noisy     & 1.97 & 3.35 & 2.44 & 2.63 & -         \\ \hhline{|=|=|=|=|=|=|}
Weiner filtering    & 2.22 & 3.23 & 2.68 & 2.67 & -         \\ \hline
Deep Feature Loss~\cite{germain2018speech}    & 2.57 & - & - & - & -         \\ \hline
Hi-Fi GAN~\cite{su2020hifi} & 2.94 & 4.07 & 3.07 & 3.49 & -         \\ \hline
Self-Adapt MHSA~\cite{koizumi2020speech}      & 2.99 & 4.15 & 3.46 & 3.51 & -         \\ \hline
T-GSA~\cite{kim2020t}     & 3.06 & 4.18 & 3.59 & 3.62 & -         \\ \hline
DEMUCS~\cite{defossez2020real}    & 3.07 & 4.31 & 3.4  & 3.63 & \textbf{95}        \\ \hline
\begin{tabular}[c]{@{}c@{}}\textbf{PERL-AE} (ours)\\ ($l_1$+AcousticEvent)\end{tabular}   & \textbf{3.17} & \textbf{4.43} & \textbf{3.53} & \textbf{3.83} & \textbf{95}        \\ \hline
\end{tabular}
}
\label{tab:sota}
\vspace{-2mm}
\end{table}

\begin{table}[t]
\centering
\caption{Reproducing ablation study of \cite{gulati2020conformer} by removing features from Conformer network to converge to vanilla Transformer.}
\resizebox{0.48\textwidth}{!}{
\begin{tabular}{|l|c|c|c|c|c|}
\hline
          & PESQ & CSIG & CBAK & COVL & STOI \\ \hline
Full Conformer & \textbf{3.17} & \textbf{4.43} & \textbf{3.53} & \textbf{3.83} & \textbf{95}        \\ \hhline{|=|=|=|=|=|=|}
- SWISH + ReLU & 3.15 & 4.38 & 3.52 & 3.78 & 94.9\\ \hline
\hspace{0.75mm}- Convolution Block & 2.81 & 4.21 & 3.36 & 3.51 & 94.7\\ \hline
\hspace{1.5mm}- Macaron FFN & 2.74 & 4.14 & 3.31 & 3.44 & 94.8\\ \hline
\hspace{2.25mm}- Relative Pos. Emb. & 2.65 & 4.01 & 3.30 & 3.36 & 94.4\\ \hline
\end{tabular}
}
\label{tab:ablation}
\end{table}

\begin{table}[t]
\centering
\caption{Comparison of fixed hand-tuned weights and multi-task weights learning methods for training PERL model (seven losses)}
\resizebox{0.48\textwidth}{!}{
\begin{tabular}{|p{2.1cm}|c|c|c|c|c|}
\hline
                         & PESQ & CSIG & CBAK & COVL & STOI \\ \hline
Hand Tuning (\textbf{PERL})             & 2.89 & 4.11 & 3.41 & 3.5  & 94.6 \\ \hline
\begin{tabular}[c]{@{}l@{}}Fine-Tuning on\\$l_1$+AcousticEvent\end{tabular}              &   3.12   &  4.4    & 3.51     & 3.78     &   95   \\ \hhline{|=|=|=|=|=|=|}
Equal weights            & 2.74 & 4.02 & 3.31 & 3.37 & 94.3 \\ \hline
Coefficient of Variation~\cite{groenendijk2020multi} & 2.74 & 4.08 & 3.31 & 3.31 & 94.3 \\ \hline
Uncertainty Weighting~\cite{kendall2018multi}    & \textbf{2.85} & 4.12 & \textbf{3.35} & \textbf{3.48} & \textbf{94.5} \\ \hline
Dynamic Weight Average~\cite{liu2019end} & 2.8  & \textbf{4.14} & 3.34 & 3.47 & 94.3 \\ \hline
GradCosine~\cite{du2018adapting}               &  2.54    &   3.92   &    3.15  & 3.22     &  93.1    \\ \hline
\end{tabular}
}
\label{tab:mtl}
\vspace{-3mm}
\end{table} 

\vspace{-3mm}
\subsection{Automatic weight learning versus hand-tuning of weights}
\label{sec:multi}
PERL faces challenges of MTL. For e.g., loss terms can have different dynamic range, speed of convergence (for fixed learning rate), and complementarity with main loss.
In this section, we investigate if \ac{SOTA} automatic weight learning methods can outperform hand-tuned PERL system i.e. the system trained with all seven losses.
We also want to investigate if detrimental losses can be automatically rejected, especially by GradCosine~\cite{du2018adapting}.
In Table \ref{tab:mtl}, second row refers to PERL system fine-tuned with the best loss discovered in Table \ref{tab:comb} i.e. ``$l_1$+AcousticEvent''.
Note that it is able to (almost) recover fully.
Results with equal weighting shows that the choice of weights is paramount.
We find GradCosine to be the worst and Uncertainty Weighting to be the best among four \ac{MTL} methods.
Our important finding is that learning weights or even dynamically adjusting (non-learnable) weights cannot outperform well-tuned (by hand) results.
We suspect this because of two reasons.
One, the domain mismatch of enhancement training data with auxiliary network data is hard to overcome.
By leveraging more training data and learning a transformation layer between auxiliary networks and the main (denoising) network, this problem can perhaps be quelled.
Two, the fundamental assumption of multi-task learning of \emph{complementarity of losses} does not hold and even SOTA MTL methods cannot overcome it by, for example, (soft/hard) rejection of terms unaligned with \emph{main loss}.

\vspace{2mm}
\section{Conclusion}
To improve perceptual metrics on small-scale speech denoising, we propose a framework called Perceptual Ensemble Regularization Loss (PERL).
PERL leverages an ensemble of variety of open-source large-scale pre-trained speech models for deriving perceptual loss or Deep Feature Loss.
We show the efficacy of all models individually as well as in certain combinations.
Acoustic event and self-supervised PASE+ models are found to be most effective.
We also find combining regular feature-domain enhancement loss to be complementary.
With an ablation study, we highlight the critical role of components of Conformer Transformer.
When using all components of PERL, we find that our seven-loss framework suffers from the challenges of Multi-Task Learning (MTL) and we suffer a performance loss.
To recover, we experiment with various MTL methods to conclude that state-of-the-art MTL methods cannot outperform greedy hand-tuning based weight selection but, interestingly, this system can be fine-tuned on the best loss ($l_1$+AcousticEvent) to restore lost performance.
In future, we can (1) build towards universal enhancement by analyzing PERL on a richer test setup, (2) incorporate automatic speech quality quantification networks like SESQA~\cite{serra2020sesqa}, and (3) learn bridges/adaptors between main and auxiliary networks to alleviate domain mismatch.

\clearpage
\bibliographystyle{IEEEbib}
\bibliography{main}

\end{document}